\documentstyle[12pt]{article}
\begin{document}
\renewcommand{\thefootnote}{\fnsymbol{footnote}}
\newpage
\pagestyle{empty}
\setcounter{page}{0}
\renewcommand{\thesection}{\arabic{section}}
\renewcommand{\theequation}{\thesection.\arabic{equation}}
\newcommand{\sect}[1]{\setcounter{equation}{0}\section{#1}}
\newfont{\twelvemsb}{msbm10 scaled\magstep1}
\newfont{\eightmsb}{msbm8}
\newfont{\sixmsb}{msbm6}
\newfam\msbfam
\textfont\msbfam=\twelvemsb
\scriptfont\msbfam=\eightmsb
\scriptscriptfont\msbfam=\sixmsb
\catcode`\@=11
\def\Bbb{\ifmmode\let\next\Bbb@\else
  \def\next{\errmessage{Use \string\Bbb\space only in math mode}}\fi\next}
\def\Bbb@#1{{\Bbb@@{#1}}}
\def\Bbb@@#1{\fam\msbfam#1}
\newfont{\twelvegoth}{eufm10 scaled\magstep1}
\newfont{\tengoth}{eufm10}
\newfont{\eightgoth}{eufm8}
\newfont{\sixgoth}{eufm6}
\newfam\gothfam
\textfont\gothfam=\twelvegoth
\scriptfont\gothfam=\eightgoth
\scriptscriptfont\gothfam=\sixgoth
\def\frak{\frak@}
\def\frak@#1{{\fam\gothfam{{#1}}}}
\def\frak@@#1{\fam\gothfam#1}
\catcode`@=12
%
%
%
\def\CC{{\Bbb C}}
\def\NN{{\Bbb N}}
\def\QQ{{\Bbb Q}}
\def\RR{{\Bbb R}}
\def\ZZ{{\Bbb Z}}
\def\cA{{\cal A}}          \def\cB{{\cal B}}          \def\cC{{\cal C}}
\def\cD{{\cal D}}          \def\cE{{\cal E}}          \def\cF{{\cal F}}
\def\cG{{\cal G}}          \def\cH{{\cal H}}          \def\cI{{\cal I}}
\def\cJ{{\cal J}}          \def\cK{{\cal K}}          \def\cL{{\cal L}} 
\def\cM{{\cal M}}          \def\cN{{\cal N}}          \def\cO{{\cal O}}
\def\cP{{\cal P}}          \def\cQ{{\cal Q}}          \def\cR{{\cal R}} 
\def\cS{{\cal S}}          \def\cT{{\cal T}}          \def\cU{{\cal U}}
\def\cV{{\cal V}}          \def\cW{{\cal W}}          \def\cX{{\cal X}}
\def\cY{{\cal Y}}          \def\cZ{{\cal Z}}
\def\qed{\hfill \rule{5pt}{5pt}}
\def\id{\mbox{id}}
\def\ggo{{\frak g}_{\bar 0}}
\def\uqggo{\cU_q({\frak g}_{\bar 0})}
\def\uqggp{\cU_q({\frak g}_+)}
\def\half{\frac{1}{2}}
\def\btf{\bigtriangleup}
\newtheorem{lemma}{Lemma}
\newtheorem{prop}{Proposition}
\newtheorem{theo}{Theorem}
\newtheorem{Defi}{Definition}
\rightline{S709.0299}

\vfill
\vfill
\begin{center}

{\LARGE {\bf {\sf 
Nonselfdual solutions for gauge fields in Schwarzschild and deSitter backgrounds for
dimension $ d\geq 4$. }}} \\[0.8cm]

{\large  A.Chakrabarti\footnote{chakra@cpht.polytechnique.fr}}
\begin{center}
{\em 

Centre de Physique Th\'eorique\footnote{Laboratoire Propre 
du CNRS UMR 7644}, Ecole Polytechnique, 91128 Palaiseau Cedex, France.\\}

\end{center}

\end{center}

\smallskip

\smallskip 

\smallskip

\smallskip

\smallskip

\smallskip 

\begin{abstract}
A particularly simple class of nonselfdual solutions are obtained for gauge fields in
Schwarzschild and deSitter backgrounds. For Lorentz signature these have finite energy and
finite action for Euclidean signature. In each case one obtains either real or a pair of
complex conjugate solutions. The actions are easily computed for any dimension $d$.
Numerical values are given for $d= 4,6,7,8,9,10$. It is explained why $d=5$ is a very
special case. Possible continuations and generalizations of the results obtained are
indicated. A particular solution for $AdS_4$ background is presented in the Appendix.
\end{abstract}

\vfill
\newpage 

\pagestyle{plain}

\sect{Introduction :}
In a previous paper $\lbrack 1 \rbrack$, almost twenty years ago, we presented very
simple,nonselfdual solutions for gauge fields in Schwarzschild and deSitter backgrounds in
four dimensions. They were further discussed in Refs. $2$ and $3$. Here they are generalized
to all dimensions $d\geq 4$, with Lorentz or Euclidean signature and $(d-1)$ spatial
dimensions. A solution in anti-deSitter background, restrited to $d=4$, is studied in the
Appendix. 

  Let us briefly recapitulate our previous results. The necessary steps for higher
dimensions, as will be seen in the next section, will then be straightforward.

  The background metrics are, for $d=4$, in standard notations,
\begin{eqnarray}
ds^2 = \mp N dt^2 + N^{-1} dr^2 + r^2 d\Omega_2 ,
\end{eqnarray}
where respectively,
\begin{eqnarray}
 N = \Bigl( 1 - \frac{2M}{r} \Big) \quad\ (Schwarzschild), 
\end{eqnarray}
and
\begin{eqnarray} 
  N = \Bigl( 1 - \Lambda r^2 \Bigr)  \quad\ (deSitter)
\end{eqnarray}
The  $SU(2)$ generators being, in terms of the Pauli matrices

$$\sigma_{ij} = \epsilon_{ijk} \Bigl(\frac{1}{2} \sigma_k \Bigr) $$
the gauge fields, for our static, spherically symmetic ansatz, are given by
\begin{eqnarray}
A_0 &&=0 {\nonumber}\\
 A_i &&= r^{-2} \Bigl( K(r) - 1 \Bigr) \sigma_{ij} x_j
  = r^{-2} \Bigl( K(r) - 1 \Bigr)\epsilon_{ijk} \Bigl(\frac{1}{2} \sigma_k \Bigr) x_j
\end{eqnarray}

Finite energy (action) is obtained for Lorentz (Euclidean) signature, corresponding to
(1.2) and (1.3) respectively, for
\begin{eqnarray}
K= \frac {r+2Ma}{r+2Mb}, \quad\ (a= -2.366, b =4.098) 
\end{eqnarray}
and
\begin{eqnarray}
K= \frac {1 + a\Lambda r^2}{1 + b\Lambda r^2}, \quad\ (a= \pm i1.732,, b = - 0.857 \mp
i0.742) 
\end{eqnarray}

The exact values of $a$ and $b$ can be found in Ref.1. ( The simple equations determining
such parameters are given in Sec.2. )

For $(1.2)$, when evaluating the energy or the action, apart from the standard angular
integration, giving a factor $4\pi$, the radial integral is over the domain $[ 2M, \infty]$.
For Lorentz signature one obtains a finite energy for this {\it outer} region. For
Euclidean signature time becomes {\it periodic} with a period $8\pi M$ and the domain $[
2M, \infty]$ for $r$ corresponds (as explained after $(1.14)$) to the {\it entire } one
covered by the Kruskal coordinates. In this context one obtains a finite {\it action} for
$(1.5)$. This action is {\it less than }$8\pi^2$. It is
\begin{eqnarray}
S=8\pi^2 (0.959) 
\end{eqnarray}

This inequality is emphasized for the following reason. Selfdual and antiselfdual solutions
can be obtained directly from the spin connectioˆns $\lbrack 1 \rbrack$. Setting
\begin{eqnarray}
A_0 &&=\pm \Bigl(\frac{1}{2} \frac{dN}{dr}\Bigr)\Bigl(\frac{1}{2}\sigma_i x_i \Bigr)
{\nonumber}\\
 A_i&&= r^{-2} \Bigl( N^{\frac{1}{2}} -1\Bigr)\Bigl(\frac{1}{2}\epsilon_{ijk}\sigma_k\Bigr)
x_j
\end{eqnarray}
one obtains the fundamental selfdual and antiselfdual solutions for the upper and the lower
sign respectively. The Euclidean action and the topological index are respectively
$$ S=8\pi^2$$
and
\begin{eqnarray}
P=\pm 1
\end{eqnarray}

{\it Thus we have a nonselfdual solution with lower action than the (anti)instanton with
the lowest nontrivial index.} (See the relevant remarks in Sec.4.) Note that for $M=0$ in
$(1.5)$, $K=1$ and hence $A_\mu =0$ in $(1.4)$. Thus the flat space limit is trivial.{\it 
The curved metric is intrinsically necessary for the existence of such a solution}.

For the deSitter case the domain of $r$ is the {\it inner} region upto the horizon, namely,
\begin{eqnarray}
\Bigl[ 0, \Lambda^{-\frac{1}{2}} \Bigr]
\end{eqnarray}
For Euclidean signature the time period is $ 2\pi\Lambda^{-\frac{1}{2}}$. The
complex action is, for upper and lower sign in $(1.6)$ respectively,

\begin{eqnarray}
S =8\pi^2 \Bigl( 1.755 \mp i4.197 \Bigr)
\end{eqnarray}

Such a complex solution can be considered in the context of a complexified gauge group,
$Sl(2,C)$ for our case. See the remaks and the references in $\lbrack 2 \rbrack$. The
possible role of complex saddle points was also discussed in $\lbrack 2 \rbrack$.

Before introducing the ansatz for gaugefields in higher dimensions (Sec.2), let us
recapitulate briefly some known but directly relevant results concerning the chosen
metrics and the traces of the generators of the gauge group $SO(d-1)$.

The Schwarzschild metric for $d\geq 4$ in spherical coordinates $\lbrack 4 \rbrack$ is
given by

 \begin{eqnarray}
ds^2 = \mp N dt^2 + N^{-1} dr^2 + r^2 d\Omega_{d-2} ,
\end{eqnarray}

where ,
\begin{eqnarray}
 N = \Bigl( 1 - \Bigl(\frac{C}{r}\Bigr)^{(d-3)} \Big) \quad\ , 
\end{eqnarray}
and $d\Omega_{( d-2)}$ is the line element on the unit $(d-2)$-sphere.

Considering directly the Euclidean signature, the Kruskal coordinates are defined by
$$e^{(2kr^*)} = \frac {1}{4} \Bigl( \eta^2 + \zeta^2 \Bigr)$$
 \begin{eqnarray}
 e^{ikt} = \Bigl( \frac{\eta -i\zeta}{\eta + i\zeta}\Bigr)^\frac {1}{2}
\end{eqnarray}
where 
$$ r^* = \int N^{-1}dr = r + C^{(d-3)} \int {\Bigl( r^{(d-3)} - C^{(d-3)}\Bigr)}^{-1}dr$$

Using

 \begin{eqnarray}
\frac {1} {\Bigl( x^n -1\Bigr)} = \frac{1}{n} \Biggl(\frac {1}{x-1} -\frac
{x^{n-2}+2x^{n-3}+\dots +(n-2)x+(n-1)}{x^{n-1}+x^{n-2}+\dots +x+1}\Biggr)
\end{eqnarray}
it is easily seen, without evaluating $r^*$ completely in terms of partial fractions, that
 \begin{eqnarray}
r^* = r + \frac{C}{(d-3)} \int {\frac{dr}{r-C}} +h(r)
 \end{eqnarray}
where the function $h(r)$ plays no role concerning the singularity at the horizon at $r=C$.
This can be verified by consructing $h(r)$ explicitly.

Thus
 \begin{eqnarray}
ds^2 = N{\Bigl(4k^2e^{2kr^*}\Bigr)}^{-1} \Bigl( d\zeta^2 +d\eta^2\Bigr)+ r^2 d\Omega_{(
d-2)} ,
\end{eqnarray}
where
$$ e^{-2kr^*}= e^{-2kr} \Bigl( r-C\Bigr)^{-\Bigl(\frac{2kC}{d-3}\Bigr)}e^{-2kh(r)}$$
hence, setting
\begin{eqnarray}
k=-\frac{(d-3)}{2C}
\end{eqnarray}
desingularizes the horizon. The domain $\lbrack C,\infty \rbrack$ ensures the positive
definiteness of $\Bigl( \eta^2 + \zeta^2 \Bigr)$ in $(1.14)$.

From $(1.14)$ the period is now found to be
\begin{eqnarray}
P_{(d)}= 2\pi{\arrowvert k\arrowvert}^{-1} = 4\pi C /{(d-3)}
\end{eqnarray}
 For $d=4$ and $C=2M$ one gets back the well known result
$$P_{(4)}=8\pi M$$
For the deSitter case, for all $d$,
$$N=\Bigl(1-\Lambda r^2\Bigr)$$
and the period remains
\begin{eqnarray}
\tilde{P}=2\pi \Lambda^{-\frac{1}{2}}
\end{eqnarray}

Another necessary ingredient in evaluating the actions of our solutions will be seen ( in
Sec.3) to be the traces of the generators of $SO(d-1)$. This will, of course, depend on the
representation chosen. We will not always specify it in the following. But let us note here
briefly the results for spinorial constructions of these generators.

Let , for $i\neq j$,
\begin{eqnarray}
L^{(n)}_{ij}= -\frac{i}{2} \gamma^{(n)}_i\gamma^{(n)}_j 
\end{eqnarray}

where $\gamma^{(n)}_i$ are the $\gamma$-matrices for $n$ spatial dimensions, satisfying the
Clifford algebra. Then the $L$'s satisfy the $SO(n)$ algebra with Hermitian convention.
Starting with the Pauli matrices for $n=3$ and
\begin{eqnarray}
L^{(3)}_{ij}= -\frac{i}{2} \sigma_i\sigma_j 
\end{eqnarray}
one proceeds in alternate steps for even and odd dimensions as follows.

Set
\begin{eqnarray}
\gamma^{(2p)}_i = \sigma_2 \otimes \gamma^{(2p -1)}_i  \quad (i=1,\dots ,2p -1)
\end{eqnarray}
and
\begin{eqnarray}
\gamma^{(2p)}_{2p} = \sigma_1 \otimes I_{(2p - 2)}
\end{eqnarray}

In the next step, one sets, for $n=(2p+1)$,
$$\gamma^{(2p+1)}_i = \gamma^{(2p)}_i  \quad (i=1,\dots ,2p )$$
\begin{eqnarray}
\gamma^{(2p+1)}_{2p+1} = \gamma^{(2p)}_1\gamma^{(2p)}_2 \dots \gamma^{(2p)}_{2p} \equiv
\Gamma^{(2p)}_{2p+1}
\end{eqnarray}
 Here $\Gamma^{(2p)}_{2p+1}$ is the generalization of the chiral matrix $\gamma_5$ for
$n=4$.

For even $n$ one can take the chiral projections
\begin{eqnarray}
\frac{1}{2}\Bigl( 1\pm \Gamma^{(2p)}_{2p+1} \Bigr) L^{(2p)}_{ij}
\end{eqnarray}
which reduces the dimension by a factor $2$.

For such spinorial constructions,
\begin{eqnarray}
TrL^{(n)}_{ij}L^{(n)}_{i^{\prime}  j^{\prime} }= \lambda_{(n)}\delta_{ii^{\prime}}
\delta_{jj^{\prime}}
\end{eqnarray}
where (ordering indices, say, as $i<j$) 
for odd $n$,
$$\lambda_{(n)} = 2^{(n -5)/2}   $$  
and for even $n$, 
$$\lambda_{(n)} = 2^{(n -4)/2} $$  

Chiral projections give, for even $n$,
\begin{eqnarray}
\lambda_{(n)} = 2^{(n -6)/2} 
\end{eqnarray}

\section{Metrics, ansatz for gauge fields and a class of solutions for
$d\geq4$:}\setcounter{equation}{0}

We consider spherically symmetric, static metrics for dimensions $d\geq 4$. For the ansatz
to be introduced below, the Kerr - Schild form of the metrics turn out to be convenient,
for computation, to start with. The complications due to the nondiagonal form will be amply
compensated by other good properties. In this form one has
\begin{eqnarray}
g_{\mu\nu}=\eta_{\mu\nu}+l_{\mu} l_{\nu}, \quad g^{\mu\nu}=\eta^ {\mu\nu}-l^{\mu} l^{\nu}
\end{eqnarray}
where

$$\eta_{00} = \eta^{00} = -1, \eta_{ij} =\eta^{ij} = \delta_{ij}$$
and

\begin{eqnarray}
l^{\mu}l^{\nu}\eta_{\mu\nu} = l^{\mu}l^{\nu}g_{\mu\nu} =0
\end{eqnarray}

For static spherical symmetry $l_0$ is a function of $r$ only, where
$$r^2= \sum_{i=1}^{d-1} x_i^2$$
and
$$l_i=l_0 \frac{x_i}{r}  \quad (i=1,2, \dots,d-1)$$
satisfying
\begin{eqnarray}
-l_0^2 +\sum_i  l_i^2 =0
\end{eqnarray}

For Schwarzschild metric in $d$ dimensios

\begin{eqnarray}
l_0^2 = (C/r)^{d-3} \qquad (C>0)
\end{eqnarray}
and for deSitter metric,for all $d$,

\begin{eqnarray}
l_0^2 = \Lambda r^2 \qquad (\Lambda >0)
\end{eqnarray}

In this paper, seeking simplicity, we present explicit solutions for these two cases only.
The Reissner-Nordstrom case with

\begin{eqnarray}
l_0^2 = (C/r)^{d-3} -(D/r^2)^{d-3}
\end{eqnarray}
will be excluded.
The standard form in spherical coordinates is given by

\begin{eqnarray}
ds^2 = - N dt^2 + N^{-1} dr^2 + r^2 d\Omega_{(d-2)} ,
\end{eqnarray}
where $ d\Omega_{(d-2)}$ is the line element on the unit $(d-2)$-sphere and

\begin{eqnarray}
N=\Bigl( 1-l_0^2 \Bigr)
\end{eqnarray}
 (For $d=4$, for example, $C=2M$ and $D=P^2 +Q^2$ give back the well-known forms.)
The coordinate transformation relating $(2.1)$ an $(2.7)$, namely
\begin{eqnarray}
x_0 = t +\int {\frac{dr}{N}} - r
\end{eqnarray}
does not affect our particularly simple ansatz for the gauge potentials to follow (with
$A_t = A_r =0$).

{\it After} constructing the solutions using $(2.1)$ the passage to Euclidean signature is
best considered ( rather than introducing imaginary $l_0$) by directly starting from
$(2.7)$, leading to 
\begin{eqnarray}
ds^2 =  N dt^2 + N^{-1} dr^2 + r^2 d\Omega_{(d-2)} 
\end{eqnarray}

The ansatz for the gauge potentials is
\begin{eqnarray}
A_0 &&=0 {\nonumber}\\
A_i &&= r^{-2} \Bigl(K(r)- 1\Bigr)L_{ij}x_j  \quad (i=1,2, \dots,d-1)
\end{eqnarray}

Here $L_{ij} =-L_{ji}$ are hermitian $SO(d-1)$ rotation matrices satisfying
\begin{eqnarray}
[L_{ij},L_{i^{\prime}j^{\prime}}]=i\Bigl(
\delta_{ii^{\prime}}L_{jj^{\prime}}+\delta_{jj^{\prime}}L_{ii^{\prime}}
-\delta_{ij^{\prime}}L_{ji^{\prime}} -\delta_{ji^{\prime}}L_{ij^{\prime}} \Bigr)
\end{eqnarray}

For constructing solutions the Lie algebra is sufficient. But for evaluating actions one
must specify the chosen representation before computing traces. Of particular interest are
the cases discussed in Sec.1. But one can implement other representations, if so desired.

Defining
\begin{eqnarray}
F_{\mu\nu}=\partial_{\mu}A_{\nu}-\partial_{\nu}A_{\mu}+i[A_{\mu},A_{\nu}]
\end{eqnarray}
one obtains
\begin{eqnarray}
F_{0i}&&=0 {\nonumber}\\
F_{ij}&&=-r^{-2}\Bigl( K^2 -1\Bigr)L_{ij}+r^{-4}\Bigl( r\frac{dK}{dr}-(K^2 -1)\Bigr)\Bigl(
x_i(L_{jk}x_k)-x_j(L_{ik}x_k)\Bigr)
\end{eqnarray}
Now using (2.1),
$$F^{0i}=g^{0\mu}g^{i\nu}F_{\mu\nu}=r^{-1}l_0^2\frac{dK}{dr}(L_{ik}x_k)$$
and
\begin{eqnarray}
F^{ij}=g^{i\mu}g^{j\nu}F_{\mu\nu}=W_1 L_{ij}+W_2 \Bigl(x_i(L_{jk}x_k)-x_j(L_{ik}x_k)\Bigr)
\end{eqnarray}
where
$$W_1=-r^{-2}\Bigl( K^2 -1\Bigr)$$
and
\begin{eqnarray}
W_2=r^{-4}\Bigl((1-l_0^2) r\frac{dK}{dr}-(K^2 -1)\Bigr)
\end{eqnarray}
The Euler-Lagrange equations of motion are, since $\mid g\mid =1$ for the KS metric,

\begin{eqnarray}
D_{\mu}F^{\mu\nu}=\partial_{\mu}F^{\mu\nu}+i[A_{\mu},F^{\mu\nu}] =0
\end{eqnarray}
It can be shown that
$$
D_{\mu}F^{\mu 0}=0
$$
and
\begin{eqnarray}
D_{\mu}F^{\mu j}=r^{-4}\Bigl( r^{-(d-6)}\frac{d}{dr} (Nr^{(d-4)}\frac{dK}{dr})-(d-3)K(K^2
-1)\Bigr)(L_{jk}x_k)
\end{eqnarray}

Hence the equations of motion reduce to a {\it single} constraint (with $N=1-l_0^2$),
\begin{eqnarray}
\frac{d}{dr} (Nr^{(d-4)}\frac{dK}{dr})=(d-3)r^{(d-6)}K(K^2-1)
\end{eqnarray}

The factor $(d-3)$ on the right corresponds to the fact that for $d=3$ one has the Abelian
case with only $L_{12}$. We exclude this by setting $d\geq 4$. The result $(2.19)$ is
obtained less simply by using directly spherical coordinates and $(2.7)$. The equation
$(2.19)$ does not satisfy the Painleve criterion [5]. Though we cannot provide a general
analysis, surprisingly simple particular solutions have been found. Interesting solutions
can be found ,{\it not for flat space} with $N=1$, but remarkably for the curved ones given
by $(2.4)$ and $(2.5)$. The special features arising for $d=5$ will be discussed below. 
  
\subsection{Case 1. Schwarzschild background:}
For 
$$N=\Bigl( 1-(C/r)^{d-3} \Bigr)$$
set 
\begin{eqnarray}
K=\frac{r^{(d-3)}+aC^{(d-3)}}{r^{(d-3)}+bC^{(d-3)}}
\end{eqnarray}

Inserting in $(2.19)$ one obtains
\begin{eqnarray}
 &&3a+b(2d-7)+(d-1)=0{\nonumber}\\
 &&a(a+b)-(d-5)b=0
\end{eqnarray}
This involves essentially solving only a quadratic in $a$ or $b$. For $d=4$ our old results
[1] are reproduced. Of the two real solutions only the one with
\begin{equation}
a=-2.366, b=4.098
\end{equation}
gives finite energy (action) for Lorentz (Euclidean) signature.

For $d=5$ one has a very special case, as is evident from the second equation of $(2.21)$.
Now the solution
\begin{equation}
a=0, b= -(4/3)
\end{equation}
leads to a divergent action since the domain of $(r/C)$ is $[1,\infty]$ and this includes a
zero of the denominator of $K$. But now one can also consider the flat limit as follows.

Setting, for $(d=5)$,
\begin{equation}
a=\hat{a}/C^2, b=\hat{b}/C^2
\end{equation}
the set $(2.21)$ reduces to
$$3(\hat{a}+\hat{b})+4C^2=0$$
\begin{equation}
\hat{a}(\hat{a}+\hat{b})=0
\end{equation}
Hence as $C\rightarrow 0$, setting ($\delta$ being an arbitrary real number)
\begin{equation}
-\hat{a}=\hat{b}=\delta^2
\end{equation}
one obtains
\begin{equation}
(K-1)=-\frac{2\delta^2}{r^2+\delta^2}
\end{equation}
Substituting this in $(2.11)$ it is seen that for the convention, say,
$$\epsilon_{1234}=1,\quad  L_{12}=-L_{34} \quad (cyclic)$$
for the chirally projected $2\times 2$ $SO(4)$ generators one obtains the famous BPST
selfdual solution in $d=4$ as a static one in $d=5$ via our limiting process. Another
convention gives the antiselfdual form.

From $d=6$ onwards the solutions become {\it complex}.The corresponding finite complex
action ,or energy, will be obtained in the follwing section. The exact values can be
obtained, for any $d$, immediately from $(2.21)$. Some numerical values, giving a direct
idea of variation with $d$, are  given below. Both upper or both lower signs are to be
taken.For
\begin{eqnarray}
&&d=6:\quad a=0.500\pm i1.500,\qquad  b=-1.300\mp i0.900 {\nonumber}\\
&&d=7:\quad a=0\pm i1.732,,\qquad b=0.857\mp i0.742 {\nonumber}\\
&&d=8: \quad a=-0.167\pm i1.863,\qquad  b= -0.722\mp i0.621 {\nonumber}\\
&&d=9: \quad a=-0.250\pm i1.984,\qquad  b=-0.659\mp i0.541 {\nonumber}\\
&&d=10: \quad a=-0.300\pm i2.100,\qquad b=-0.623\mp i0.485 
\end{eqnarray}
\subsection {Case 2. deSitter background:}

For
$$N=(1-\Lambda r^2)$$
one can satisfy $(2.19)$ by setting 
\begin{equation}
K=\frac{1+a\Lambda r^2}{1+b\Lambda r^2}
\end{equation}
with
\begin{eqnarray}
a(a+b)+2(d-5)b=0{\nonumber}\\
3(d-3)a-(d-11)b+2(d-1)=0
\end{eqnarray}
Exact solutions can again be obtained by solving a quadratic in $a$ or $b$. Approximate
numerical values are presented below.

For $d=4$ our old results [1,2] are reproduced with
\begin{equation}
a=\pm i1.732, b=-0.857 \mp i0.742
\end{equation}

For the special case $d=5$ the eqations reduce to
\begin{eqnarray}
a(a+b)=0{\nonumber}\\
3(a+b)+4=0
\end{eqnarray} 
The one consistent solution
\begin{equation}
a=0, b=-(4/3)
\end{equation}
leads to divergence in $K$ since (see Sec.3) one considers the domain $0\leq \Lambda r^2
\leq 1$. A flat space limit can again be considered as for the Schwarzschild case. This
needs no repetition.

From $d=6$ onwards ( exhibiting a behavior complemetary to the Schwarzschild case ) the
solutions become {\it real}. From $d=6$ to $d=10$, as will be seen in Sec.3, one has two
real solutions, {\it both} giving finite action (or energy). for $a$ and $b$ one obtains
the following results.
For $d=6$:
\begin{eqnarray}
&&(1)\quad a=0.807,\quad  b=-0.547 {\nonumber}\\
&&(2)\quad a=-6.193,\quad  b=9.147 {\nonumber}
\end{eqnarray}
For $d=7$:
\begin{eqnarray}
&&(1)\quad a=-0.910,,\quad  b=-0.268 {\nonumber}\\
&&(2)\quad a=-6.589,\quad  b=16.768 {\nonumber}
\end{eqnarray}
For $d=8$:
\begin{eqnarray}
&&(1)\quad a=-0.901,\quad  b=-0.159 {\nonumber}\\
&&(2)\quad a=-7.765,\quad  b=34.159 
\end{eqnarray}
For $d=9$:
\begin{eqnarray}
&&(1)\quad a=-0.877,\quad  b=-0.108 {\nonumber}\\
&&(2)\quad a=-9.123,\quad  b=74.108{\nonumber}
\end{eqnarray}
For $d=10$:
\begin{eqnarray}
&&(1)\quad a=-0.853,\quad  b=-0.080 {\nonumber}\\
&&(2)\quad a=-10.547,\quad  b=203.480 {\nonumber}
\end{eqnarray}

In the set $(1)$ the negetive value of $b$ satisfy $\mid b\mid \leq 1$ and in the set$(2)$
$b>0$. Thus in both cases divergence will be seen to be avoided (Sec.3).
One may notice that variations (with $d$) are relatively small for  the solutions $(1)$ as
compared to the solutions $(2)$. It will be seen however (Sec.3) that the values of the
action (or energy) for the two sets remain quite close for each $d$.

For $d=11$ there is a sudden change. There is only one real solution with
\begin{equation}
a=-0.833,\quad  b=0.062
\end{equation}

If the numerical values of $(a,b)$ for some particular $d$ are are found to be of interest
in some context, they can be immediately obtained from $(2.30)$. 
    
\section{Action (energy) for Euclidean (Lorentz) signature:}\setcounter{equation}{0}

Since $\mid g\mid =1$ for $(2.1)$ and $F_{i0}=0$ for our ansatz, for both signatures one
computes, to start with,
\begin{equation}
\frac {1}{2} Tr\int dV_{(d-1)} \Bigl( F_{ij}F^{ij}\Bigr)  
\end{equation}
where $dV_{(d-1)}$ is the volume element for the spatial dimentions. For our static,
spherically symmetic ansatz the angular integrations merely give a factor equal to the
surface area of the unit $(d-2)$-sphere,namely,
\begin{equation}
\Sigma_{(d-2)}=\frac {2\pi^{(d-1)/2}}{\Gamma ((d-1)/2)}
\end{equation}
The radial integration corresponds, for Schwarzschild and deSitter backgrounds respectively,
to the domains
 \begin{equation}
[C,\infty]
\end{equation}
and
\begin{equation}
[0,\Lambda^{\frac{1}{2}}]
\end{equation}
For Lorentz signature $(3.3)$ corresponds to the {\it outer region} down to the horizon at
$r=C$. For Euclidean signature it corresponds to the {\it full domain} of reality of the
Kruskal-type coordinates (Sec.1). For both signatures $(3.4)$ corresponds to the inner
region bounded by the horizon at $r=\Lambda^{\frac {1}{2}}$.

For Lorentz signature one thus obtains a finite energy for our solutions. For Euclidean
signature the time becomes periodic, for the Schwarzschild and the deSitter metrics
respectively, with a period (Sec.1),
\begin{equation}
P_{(d)} = \frac {4\pi C}{(d-3)}
\end{equation}
\begin{equation}
\tilde{P} =\frac {2\pi}{\Lambda^{\frac{1}{2}}} 
\end{equation}
Hence by multiplying the integral $(3.1)$ by $P_{(d)}$ and $\tilde{P}$ respectively one
obtains now the total action.
The radial integral will be real or complex according to the background and the
dimension considered. For complex solutions one can obtain a real action by considering a
doubled block-diagonal form of the rotation matrices $L_{ij}$ and thus treating the complex
conjugate solutions together. Keeping various possibilities in mind, at this stage , let us
set (using the ordering $i<j$ to avoid ambiguities)
\begin{equation}
Tr(L_{ij}L_{i^{\prime}j^{\prime}})=\lambda_{(d)} \delta_{ii^{\prime}}\delta_{jj^{\prime}}
\end{equation}
(See Sec.1 for evaluation of $\lambda_{(d)}$ corresponding to spinorial constructions,
setting $d=(n+1)$.)
To evaluate $(3.1)$ we need the traces, with $(i,j=1,\dots ,d-1)$,
\begin{eqnarray}
&&T_1=Tr(L_{ij}L_{ij}) {\nonumber}\\
&&T_2=r^{-2}Tr\Bigl(L_{ij}\Bigl((x_{i}L_{jk}x_k)-(x_{j}L_{ik}x_k)\Bigr)\Bigr) \\
&&T_3=r^{-4}Tr\Bigl((x_{i}L_{jk}x_k)-(x_{j}L_{ik}x_k)\Bigr)^2 {\nonumber}
\end{eqnarray}
Careful counting leads to
\begin{eqnarray}
&&T_1=(\lambda_{(d)}/2)(d-1)(d-2){\nonumber}\\
&&T_2= -\lambda_{(d)}(d-2)\\
&&T_3=\lambda_{(d)}(d-2) {\nonumber}
\end{eqnarray}

Using $(2.14)$ and $(3.9)$ one obtains after grouping terms,
\begin{equation}
\frac{1}{2}TrF_{ij}F^{ij}=\frac{\lambda_{(d)}}{2}(d-2)r^{-4}\Bigl(2N\Bigl(r{\frac{dK}{dr}}\Bigr)^2
+(d-3){(K^2 -1)}^2\Bigr)
\end{equation}
\subsection{The radial integral:}
(1) For the {\it Schwarzschild} case the radial integral is

\begin{eqnarray}
&&\int_C^{\infty} dr r^{(d-6)}\Bigl(2N\Bigl(r\frac{dK}{dr}\Bigr)^2 +(d-3)\Bigl(K^2
-1\Bigr)^2\Bigr){\nonumber}\\
&&C^{(d-5)}\int _1^{\infty}dxx^{(d-6)}\Bigl(2N\Bigl(x\frac{dK}{dx}\Bigr)^2 +(d-3)\Bigl(K^2
-1\Bigr)^2\Bigr)\\ 
&&\equiv C^{(d-5)} I_{(d)}{\nonumber}
\end{eqnarray}
where now
$$N=\Bigl(1-x^{-(d-3)}\Bigr),\quad K=\frac{x^{(d-3)}+a}{x^{(d-3)}+b}$$
The total Euclidean action , for example, is
\begin{equation}
(1/2)P_{(d)} \Sigma _{(d-2)}\lambda_{(d)}(d-2)C^{(d-5)} I_{(d)}
\end{equation}

For $d=4$
\begin{equation}
P_{(4)}=4\pi C=8\pi M,\quad \Sigma_{(2)} =4\pi ,\quad  \lambda_{(4)} = (1/2)
\end{equation}
leading to a total action
\begin{equation}
8\pi^2 I_{(4)} = 8\pi^2 (0.959)
\end{equation}
which reproduces exactly our old result [1].

Leaving aside the exceptional case $d=5$, we give below the values of $I_{(d)}$ for the
solutions $(2.28)$. $I_{(d)}$ can be expressed in terms of standard integrals for the
general case. Direct numerical evaluation gives , corresponding to the upper and the lower
sign in $(2.28)$ respectively,
\begin{eqnarray}
&&I_{(6)}=7.605 \mp i9.141 {\nonumber}\\
&&I_{(7)}=7.021 \mp i16.788 {\nonumber}\\
&&I_{(8)}=5.217 \mp i25.378 \\
&&I_{(9)}=2.109 \mp i34.906{\nonumber}\\
&&I_{(10)}= -2.320 \mp i45.312 {\nonumber}
\end{eqnarray}
Note that the real part becomes negetive for $d=10$, indicating dominant contribution of
terms biliear in the imaginary parts.

(2) For the {\it deSitter} case the radial integral is
\begin{eqnarray}
&&\int_0^{\Lambda^{-1/2}} dr r^{(d-6)}\Bigl(2N\Bigl(r\frac{dK}{dr}\Bigr)^2 +(d-3)\Bigl(K^2
-1\Bigr)^2\Bigr){\nonumber}\\
&&=\Lambda^{-(d-5)/2}\int _0^{1}dxx^{(d-6)}\Bigl(2N\Bigl(x\frac{dK}{dx}\Bigr)^2
+(d-3)\Bigl(K^2 -1\Bigr)^2\Bigr)\\ 
&&\equiv \Lambda^{-(d-5)/2} \tilde{I}_{(d)}{\nonumber}
\end{eqnarray}
where now
$$N=\Bigl(1-x^{2}\Bigr),\quad K=\frac{1+ax^2}{1+bx^2}$$
The total Euclidean action , for example, is
\begin{equation}
(1/2) {\tilde {P}} \Sigma
_{(d-2)}\lambda_{(d)}(d-2)\Lambda^{-(d-5)/2}{\tilde{I}}_{(d)}
\end{equation}

For $d=4$,with $\tilde{P} =2\pi \Lambda^{-1/2}$, one obtains the total action as
\begin{eqnarray}
4\pi^2 \tilde{I}_{(4)} &&= 4\pi^2 (3.510 \mp i8.394) {\nonumber}\\
&&=8\pi^2 (1.755\mp i4.197) 
\end{eqnarray}
This is again our old result $[1,2]$.

Leaving aside again the case $d=5$, we give the values of $\tilde{I}_{(d)}$ for the
solutions $(2.34)$. For each value of $d$ (from $6$ to $10$) one now has two real solutions
($(1)$ and $(2)$ in $(2.34)$). Considering them in order one obtains the following values:

\begin{eqnarray}
(1)\quad  \tilde{I}_{(6)} &&=0.495 ,\qquad    (2)\quad \tilde{I}_{(6)} =2.447 {\nonumber}\\
(1)\quad \tilde{I}_{(7)} &&=1.112 ,\qquad  (2)\quad \tilde{I}_{(7)} =1.824 {\nonumber}\\
(1)\quad \tilde{I}_{(8)} &&=1.134 ,\qquad  (2)\quad \tilde{I}_{(8)} =1.589 \\
(1)\quad \tilde{I}_{(9)} &&=1.096 ,\qquad  (2)\quad \tilde{I}_{(9)} =1.473 {\nonumber}\\
(1)\quad \tilde{I}_{(10)} &&=1.056 ,\qquad  (2)\quad \tilde{I}_{(10)} =1.395 {\nonumber}
\end{eqnarray}

Finite values are obtained in each case since $(1+b\Lambda r^2)$ has no zero in the domain
$[0,1]$ of $\Lambda r^2$. For the anti-deSitter case there is no horizon and our solutions,
even apart from the question of the values of $b$, lead to divergent actions due to
asymptotic properties for $d \geq 6$.

\section{Remarks:}

\setcounter{equation}{0}

Various aspects of nonselfdual solutions have been studied by a number of authors. An
incomplete list of references is provided [6,7,8,9,10,11]. Here we have shown how the
simplest static, spherically symmetric curved spaces provide surprising new possibilities.
Our results are not limited to existence theorems. We provide explicit solutions and hence
complete information concerning the gauge potentials at every space-time point. The
simplicity of our solutions should permit a relatively easy study of normal modes and
(un)stability properties. Maintaining the spherical symmetry one can proceed as for
sphalerons [12,13,14,15,16], the best known class of nonselfdual solutions. The starting
point, without Higgs fields and with our explicit solutions for the gauge potentials,
should be even simpler. Such a study would yet involve considerable numerical computations.
It is deferred to another paper. But we add some relevant comments.

We started by pointing out the striking fact that the action $(1.7)$ of our nonselfdual
solution (Schwarzschild,$d=4$) is lower than that for the lowest-action, non-trivial,
selfdual solution $(P=1)$. But our solution is not topologically stabilized. As an evident
consequence of $F_{0i}=0$ it has zero index $(P=0)$. Considering only $F_{ij}$, it is seen
that as $r\rightarrow \infty$ the magnitude falls faster $(\approx r^{-3})$ than that for
a monopole. Presumably our solution provides a minimal saddle-point between two
topologically distinct vacua. But a precise statement needs further study.
Apart from the common factors ( $4\pi$ from angular integrations and $8\pi M$ for the
period) one can compare the radial contributions. The action densities of the selfdual and
the nonselfdual cases are respectively
\begin{equation}
\frac{6M^2}{r^4}
\end{equation}
and ( with $ a=- 2.366$, $b=4.098$ ) 
\begin{equation}
\frac{4M^2 (a-b)^2}{r^2 (r+2Mb)^4}\Bigl(3r^2+ 2M(2(a+b)-1)r + 2M^2(a+b)^2\Bigr)
\end{equation} 
As $r\rightarrow \infty$ both fall as $r_4$ but for $(4.2)$ with a larger numerical
coefficient. On the other hand $(4.2)$ starts from a lower value at $r=2M$. The total
effects are given by $(1.9)$ and $(1.7)$ respectively.
One further point should be noted. For Euclidean signature selfdual gauge field
configurations, having zero enegy-momentum tensor, do not perturb the metric. Thus one
obtains, effectively, a solution for the {\it  combined} gravitation-gauge field system. For
a nonselfdual system this is no longer the case. Hence there a "background approximation" (
ignoring the back reaction of the gauge field on the metric) is involved. But again, this
is true not only for cuved but also for a flat background, the latter being held fixed to be
flat even in presence of other fields.

For deSitter background $(d=4)$ we obtained a pair of complex solutions. The conformal
properties of the metric and of gauge fields  in four dimensions permitted us to
reinterprete our static solutions as time-dependent complex ones in flat space-time [2].
This is in sharp contrast to the Schwarzschild case where the solution vanishes with $M$.

 After the very special case of $d=5$, discussed separately, from $d=6$ onwards the
complementary behaviours for these metrics continue with a crossover. Schwarzschild
solutions become complex while the deSitter ones become real. It is remarkable that {\it
both} real solutions $(d=6,7,8,9,10)$ give finite action, the ratio of the two actions
changing slowly with $d$. The prospect of mapping out the action landscape, with ridges
and valleys, in the neighbourhoods of these solutions is intriguing. The special nature
of $d=11$ has been pointed in $(2.35)$.

The solutions of the standard Yang-Mills Lagrangian are no longer independent of
conformal factors in the metric for $d> 4$. (In particular, the lack of scale invariance
is signalled by the persistence of the parameters $C$ and $\Lambda$ in the actions.) So
a direct passage to flat space is not possible , preventing a comparison of the actions of
our real deSitter soutions with those of possible flat-space solutions in higher
dimensions -- the fundamental octonionic instanton [17,18] in eight dimensions, for
example. For a Lagrangian {\it quartic} in $F_{\mu\nu}$ conformal invariance is
restored in $8$ dimensions.( See Ref.19 and the references cited there for the general
case of $4p$ dimensions.)     But our class of nonselfdual solutions have not been
constructed in the framework of these generalized Lagrangians.

We have left various possible generalizations unexplored. To start with one can consider
more general metrics as backgrounds. Solving $(2.19)$ with $N$ given by the
Reissner-Nordstrom metric is the most immediate possibilty.( But one encounters here new
problems.) For both $d=4$ and $d=8$ the Kerr metric can be related to the Schwarzschild one
through an imaginary translation [20]. One can implement a corresponding translation in the
gauge potentials to see whether it can be adapted, or not, to obtain solutions for Kerr
backgrounds. The $AdS$ background, briefly introduced in the Appendix, evidently deserves a
more thorough study. We hope to explore such possibilities elsewhere.

One reason for presenting our solutions, restricted as they are, is the pleasant simplicity
attained. We show that curvature, in some cases, can open doors rather than erect
barriers. Another reason is the current broad interest, in the context of strings and
branes, in solutions for higher dimnsions. Without citing references, let us state that
 nonselfdual, non-BPS, solutions deserve scrutiny.

\vspace{1cm}

I thank M.Duneau, F.Nogueira, B.Pire, A.Ramani, C.Roiesnel and D.H.Tchrakian for help.

\newpage
\section*{Appendix: A nonselfdual solution for $AdS_4$ background.}
\setcounter{equation}{0}
\renewcommand{\theequation}{A.\arabic{equation}}

For $d=4$ a divergent solution in deSitter background was shown [1,3] to be related through
conformal transformations to meron-type solutions in flat space-time. This, particularly
simple, solution corresponds (with $\Lambda >0$) to
\begin{equation}
N=(1-\Lambda r^2), \quad K=N^{-\frac{1}{2}}=(1-\Lambda r^2)^{-\frac{1}{2}}
\end{equation}
Here we note that changing the sign before $\Lambda$ one obtains the anti-deSitter case
($AdS_4$)  and
\begin{equation}
N=(1+\Lambda r^2), \quad K=N^{-\frac{1}{2}}=(1+\Lambda r^2)^{-\frac{1}{2}}
\end{equation}
This provides a solution of $(2.19)$ with $d=4$ namely of
\begin{equation}
\frac{d}{dr}\Bigl(N\frac{dK}{dr}\Bigr)=r^{-2}K\Bigl(K^2 -1\Bigr)
\end{equation}
{\it where now} $K$ {\it is no longer singular.} Henceforth, in the Appendix, we set
$\Lambda =1$ for simplicity.The radial integral $(3.16)$, in absence of the horizon,should
now be replaced by
\begin{equation}
\tilde{I}_{(4)}=\int_0^{\infty} dx x^{-2}\Bigl(2N\Bigl(x\frac{dK}{dx}\Bigr)^2 + \Bigl(K^2
-1\Bigr)^2 \Bigr)
\end{equation}
where
$$N=(1+x^2), \quad K=(1+x^2)^{-\frac{1}{2}}$$
One obtains
\begin{equation}
\tilde{I}_{(4)}= 3\int _0^{\infty} \frac{x^2}{(1+x^2)^2}dx =\frac{3\pi}{4}
\end{equation}
Thus ,corresponding to $(3.10)$, one obtains a finite spatial integral
 \begin{equation}
\frac{3\pi^2}{2}
\end{equation}
The factor from the time integration depends on the chosen context. Now there is no horizon
to be desingularized and the discussion of Sec.1 is not directly relevant. But one can
start by considering the hypersurface
\begin{equation}
-t_1^2-t_2^2+x_1^2+x_2^2+x_3^2=-1
\end{equation}
In terms of the spherical coordinates
\begin{eqnarray}
(x_1,x_2,x_3)&&\rightarrow (r,\theta,\phi){\nonumber}\\
(t_1,t_2)&&\rightarrow (T,\psi)
\end{eqnarray}
the metric on the hypersurface
$$r^2 - T^2 =-1$$
is
\begin{equation}
ds^2 = -(1+r^2)d\psi^2 + (1+r^2)^{-1}dr^2+ r^2 d\Omega_2
\end{equation}
In this context the $\psi$-integration gives a factor $2\pi$ and one obtains a total action
\begin{equation}
3\pi^3
\end{equation}
But often it is preferable to consider the covering space $(CAdS)$ replacing $\psi \in
S^1$ by $t\in R$. Then the action is evidently divergent.

The solution of $(A.1)$ with the square root involved for $K$ seems to be specific to $d=4$.
But it would be interesting to search for suitable generalizations, related to this class,
for higher dimensions.

\end{document}